# Wafer-scale epitaxial growth of the thickness-controllable van der Waals ferromagnet CrTe$_2$ for reliable magnetic memory applications

Xinqi Liu[1,2], Yunyouyou Xia[1,2], Lei Gao[3], Puyang Huang[4], Liyang Liao[5], Baoshan Cui[6], Dirk Backes[7], Gerrit van der Laan[7], Thorsten Hesjedal[8], Yuchen Ji[1,2], Peng Chen[1,4], Fan Wu[2], Meixiao Wang[1,2], Junwei Zhang[3], Guoqiang Yu[6,9], Cheng Song[5], Yulin Chen[1,2,8], Zhongkai Liu[1,2], Yumeng Yang[4], Yong Peng[3], Gang Li[1,2], Qi Yao[1,2]\*, and Xufeng Kou[1,4]\*

[1]ShanghaiTech Laboratory for Topological Physics, ShanghaiTech University, Shanghai 201210, China.

[2]School of Physical Science and Technology, ShanghaiTech University, Shanghai 201210, China.

[3]School of Materials and Energy, Lanzhou University, Lanzhou 730000, China.

[4]School of Information Science and Technology, ShanghaiTech University, Shanghai 201210, China.

[5]Key Lab Advanced Materials (MOE), School of Materials Science and Engineering, Beijing Innovation Center for Future Chips, Tsinghua University, Beijing 100084, China.

[6]Songshan Lake Materials Laboratory Dongguan, Guangdong 523808, China.

[7]Diamond Light Source, Harwell Science and Innovation Campus, Didcot, Oxfordshire OX11 0DE, United Kingdom.

[8]Department of Physics, University of Oxford, Oxford OX1 3PU, United Kingdom.

[9]Beijing National Laboratory for Condensed Matter Physics Institute of Physics Chinese Academy of Sciences Beijing 100190, China.

\*Correspondence to: yaoqi@shanghaitech.edu.cn; kouxf@shanghaitech.edu.cn.

**To harness the intriguing properties of two-dimensional van der Waals (vdW) ferromagnets (FMs) for versatile applications, the key challenge lies in the reliable material synthesis for scalable device production. Here, we demonstrate the epitaxial growth of single-crystalline 1$T$-CrTe$_2$ thin films on 2-inch sapphire substrates. Benefiting from the uniform surface energy of the dangling bond-free Al$_2$O$_3$(0001) surface, the layer-by-layer vdW growth mode is observed right from the initial growth stage, which warrants precise control of the sample thickness and atomically smooth surface morphology across the entire wafer. Moreover, the presence of the Coulomb interaction at the CrTe$_2$/Al$_2$O$_3$ interface serves as an effective tuning parameter to tailor the anomalous Hall response, and the structural optimization of the CrTe$_2$-based spin-orbit torque device leads to a substantial switching power reduction by 54%. Our results may lay out a general framework for the design of energy-efficient spintronics based on configurable vdW FMs.**

Featured by the inherent long-range magnetic order and a high-degree of magnetism tunability under external stimuli, the discovery of two-dimensional (2D) van der Waals (vdW) ferromagnets (FMs) has brought about a new platform for the advancement of energy-efficient non-volatile spintronics[1, 2]. In the meanwhile, the weak inter-layer vdW force would facilitate the hetero-integration between 2D FMs and other materials with different physical orders, hence greatly broadening the functionalities of the associated device applications[3, 4]. To obtain high-quality 2D FMs with aforementioned characteristics, the most frequently-adopted synthetic methods are the mechanical and electrochemical transfer techniques (*i.e.,* the so-called top-down method), where a-few-layer flakes are exfoliated from their bulk crystals[1, 2, 5]. Although it seems to be cost-effective and straightforward, this approach inevitably suffers from low productivity, small sample size, and limited thickness/orientation controllability[1, 5, 6]. To comply with the mass-production prerequisite, alternative bottom-up methods, such as the chemical vapor deposition (CVD)[7], the seeded growth[8] and dual-coupling-guided growth[9], are developed to fabricate large-size vdW FMs. However, the 2D FMs grown via chemical reaction under the equilibrium condition would follow the Wulff construction which constantly leads to the edge-induced nucleation and the formation of self-limited triangular or hexagonal islands[10-12]. Even though dedicated surface treatments have recently been introduced to permit uniform nucleation[8, 9, 11, 12], the high growth rate would make the film thickness control challenging. As a result, the layer-dependent electronic/magnetic properties of vdW FMs and the opportunities to implement their heterostructure/superlattice in spintronics remain elusive.

In this work, we report the wafer-scale growth of uniform $CrTe_2$ thin films and $Bi_2Te_3/CrTe_2$ heterostructures by molecular beam epitaxy (MBE). Owing to its low-energy physical deposition nature and non-equilibrium growth dynamics[13-15], MBE ensures high homogeneity with atomic-scale surface roughness in the grown 2-inch samples, and the vdW layer-by-layer growth mode empowers us with the ability to finely tune the film thickness across the 2D-to-bulk regions. Together with the *ab initio* calculation of electronic band structure, we unveil that the quantum confinement and interfacial effects in the ultra-thin $CrTe_2$ film would effectively modify its Berry curvature and magnetic order strength, triggering the polarity change of the resulting anomalous Hall signal. Furthermore, by combining the ferromagnetic $CrTe_2$ with the topological insulators $Bi_2Te_3$, we realize the spin-orbit torque (SOT)-driven magnetization switching in the integrated vdW heterostructures with a high endurance of more than $5 \times 10^4$ read-write cycles, and the manageable thickness of the $CrTe_2$ layer offers an additional degree of freedom for tailoring the performance of the SOT devices.

**Wafer-scale epitaxial growth of vdW FM CrTe$_2$ thin films.** Experimentally, 2-inch Al$_2$O$_3$(0001) wafers were chosen as the substrate. The absence of dangling bonds at the substrate surface and the weak bonding energy between CrTe$_2$ and Al$_2$O$_3$ guarantee a uniform surface energy distribution, which in turns promotes the van der Waals epitaxial growth of CrTe$_2$[16, 17]. Accordingly, a sharp reflection high-energy electron diffraction (RHEED) streaky pattern develops already for the first layer, and the spacing between the two first-order reciprocal lattice rods in **Fig. 1a** corresponds to an in-plane lattice constant of $a$ = 3.84 Å, in agreement with expected 1$T$-CrTe$_2$ data[18]. Further, high-resolution transmission electron microscopy (TEM) (**Fig. 1b**) shows an intact, layered crystalline structure with abrupt van der Waals gaps, and within each constituent layer, the Cr (pink spheres) and Te (yellow spheres) atoms rigorously follow the Te-Cr-Te Z-shaped stacking configuration, validating the 1$T$ phase of the as-grown CrTe$_2$ sample[7].

Next, we conducted atomic force microscope (AFM) measurements in order to examine the conformity of the CrTe$_2$ film, and **Fig. 1c** illustrates one dataset of the 9 monolayer (ML) sample with two noticeable features. First of all, in contrast to μm-size triangular or hexagonal terraces (*i.e.,* as they would stem from the island nucleation process) observed in conventional vdW 2D materials[19, 20], the AFM mapping of the CrTe$_2$ film shows quite uniform nucleation and homogeneous atomic-scale surface morphology (*i.e.,* the root-mean-square roughness $R_q$ = 3.68 Å is much less than the height of one CrTe$_2$ ML), which may take advantages of both the surface energy reduction (*i.e.,* owing to the vdW gap at the CrTe$_2$/Al$_2$O$_3$ interface) and the long diffusion length of the adatoms[11]. Secondly, the histogram of $R_q$ collected from 36 different areas exhibits a normal distribution with a narrow standard deviation of $3\sigma$ = 0.28 Å, hence manifesting the high uniformity across the whole 2-inch wafer (Supplementary Section 1).

In terms of film crystallinity, **Fig. 1d** displays the X-ray diffraction (XRD) results of five MBE-grown CrTe$_2$ samples with thicknesses ($d$) ranging from 9 to 33 ML. It can be clearly seen that in addition to the substrate signal, all spectra display only a series of CrTe$_2$ (00*n*) diffraction peaks without any impurity phase within the instrument detection limit, and the extracted out-of-plane lattice parameter $c$ = 6.10 Å is consistent with the 1*T*-CrTe$_2$ structure as well. Strikingly, **Fig. 1e** reveals identical normalized CrTe$_2$ (002) *Ω*-scan curves with negligible mosaicity broadening (*i.e.,* no misorientation of crystallites) compared with the Al$_2$O$_3$ (0006) reference line-shape, and the full width at half maximum (FWHM) of 0.037°±0.001° is among the narrowest values reported for vdW films[20-22]. Besides, the pronounced Kiessig fringes observed in the X-ray reflection (XRR) data in **Fig. 1f** not only quantify the film thickness, but also attest the layer-by-layer epitaxial growth mode and low interlayer roughness of the CrTe$_2$ samples. In conclusion, our comprehensive structural characterizations confirm the high quality and reproducibility of the wafer-scale 1*T*-CrTe$_2$ thin films, therefore offering a solid foundation for reliable device applications.

**Tunable anomalous Hall response driven by temperature and dimensionality.** To determine the magnetic properties of the epitaxial 1*T*-CrTe$_2$ thin films, we fabricated mm-scale six-probe Hall bar devices on the 2-inch wafers (**Fig. 2a**), and performed magneto-transport measurements at low temperatures. **Figure 2b** exemplifies the characteristic anomalous Hall effect (AHE) on a 14 ML CrTe$_2$ sample, where the nearly square-shaped $R_{xy}$ hysteresis loop and a butterfly-type double-split magneto-resistance (MR) curve highlight the well-established spontaneous magnetization at $T$ = 10 K. Likewise, the strong perpendicular magnetic anisotropy (PMA) of the films is also verified by the element-specific X-ray magnetic circular dichroism (XMCD) hysteresis loops recorded at the Cr $L_3$ absorption edge, confirming that the magnetism originates from the Cr atoms (Supplementary Section 2). As visualized in **Fig. 2c**, the XMCD loop exhibits the same behavior as the AHE one as a function of the applied magnetic field, which identifies that the ferromagnetism evolves together with the anisotropy of the orbital moment[23].

Given that the reduced dimensionality can help to modify the electronic band structure and magnetic properties of vdW 2D FMs[1, 2, 5, 24], we subsequently conducted systematic magneto-electric transport measurements on three CrTe$_2$ samples with $d$ = 5, 9, and 14 MLs. Remarkably, the AHE results show a distinct thickness-dependent trait. As evidenced in **Fig. 2d1-d3**, the coercive field $H_C$ decreases from 1.17 T (14 ML) to 1.15 T (9 ML) and 0.98 T (5 ML) at $T$ = 1.5 K, and this evolution trend suggests the weakening of the magnetism with reduced dimensionality. More importantly, the AHE hysteresis loop reverses its polarity from negative to positive when the thickness is reduced to 5 ML. Similarly, such a negative-to-positive transition behavior of the AHE polarity is also observed in the bulk-type 14 ML and 9 ML CrTe$_2$ samples during the warming-up process, yet the corresponding transition temperature $T^*$ is reduced from 115 to 70 K (*i.e.,* in contrast, the AHE sign remains to be positive in the 2D-limit 5 ML thin film regardless of the base temperature).

To understand this dimensionality- and temperature-induced anomalous Hall response, we applied density-functional theory (DFT) calculations to elaborate the underlying physics[25-28]. In particular, by introducing the on-site Coulomb potential $U$ correction to account for the correlation effects from the Cr $d$-orbitals[29, 30], the theoretical derived Hall conductivity $\sigma_{xy}$ is found to be positive in the monolayer CrTe$_2$ system (**Fig. 3a**) and its sign changes to negative in the bulk case (**Fig. 3b**), reproducing the observed thickness-dependent AHE polarity-reversal. To quantify this low-dimensional effect, we conducted systematic investigations by varying the Coulomb potential amplitudes in both the monolayer and the bulk CrTe$_2$ systems. As outlined in **Fig. 3c**, the negative $\sigma_{xy}$ of the bulk form (*i.e.,* with periodic boundary condition) shows little variation with respect to $U$. On the contrary, once the film thickness enters the 2D domain, the itinerant electrons in the few-layer channel are subject to a confined out-of-plane degree of freedom, and they naturally possess lower kinetic energy than those in the bulk region. Under such circumstances, the potential energy $U$ created by the Coulomb repulsion effectively increases, which thereafter modulates the Berry curvature near the Fermi level (**Fig. 3a**) and leads to a $σ_{xy}$ sign-reversal. Along with the experimental observations (**Figs. 2d1-d3**), the DFT simulations elucidate critical influence of the interface-related effect on tailoring the electronic geometry and AHE response in the 2D $CrTe_2$ system.

Likewise, we can attribute the temperature-dependence of the anomalous Hall resistance in the thick $CrTe_2$ samples (*i.e., d* ≥ 9 ML) to the modulation of Berry curvature associated with the magnetization change[28, 31]. Specifically, the diminished magnetic moment at elevated temperatures may bring about a re-distribution of electronic states and switch the AHE polarity below a critical value of $M_C$ (*i.e.*, above a characteristic transition temperature $T^*$)[31], given that the Berry curvature is highly sensitive to the occupied density of states near the Fermi level. Following the same scenario, as the overall magnetization is gradually suppressed in the thinner films, it is expected that the 9 ML $CrTe_2$ sample would reach $M_C$ at a lower transition temperature $T^*$ compared to the 15 ML counterpart. In the ultra-thin film case (*i.e., d* = 5 ML) where the saturated magnetic moment is always smaller than $M_C$, its temperature-invariant positive AHE response is therefore in line with that of the bulk-type $CrTe_2$ system in the high-temperature region.

**Energy-Efficient SOT switching in $Bi_2Te_3$/$CrTe_2$ heterostructures.** Spin-orbit-torque (SOT) has been utilized for next-generation magnetic random-access memory (MRAM) technology, owing to its low power consumption and logic-programmable capability[32, 33]. Considering that the SOT-driven magnetization switching is determined by the charge-to-spin conversion efficiency, it has been discovered that topological quantum materials (*e.g.*, topological insulators, Weyl/Dirac semimetals), in which the intrinsic spin-orbit coupling (SOC) is strong enough to cause band inversion, can enable sufficient current-induced spin polarization along the surface/edge states through the spin-momentum locking mechanism[34, 35]. Apart from the spin current channel, the magnetic strength in the adjacent FM layer is another key factor to affect the switching current level. In this regard, the tunable magnetism identified in our low-dimensional $CrTe_2$ samples, assuming that they can be incorporated into a SOT structure, could afford more flexibility for optimizing the device performance.

Accordingly, to fully explore the potential of $CrTe_2$ for SOT-MRAM applications, we adopted the same vdW growth procedure to prepare $Bi_2Te_3$/$CrTe_2$ heterostructures by MBE (*i.e.,* in order to make a fair comparison, the top $Bi_2Te_3$ layer thickness is fixed to 18 nm in all samples investigated in this work). As shown in **Fig. 4a**, the ultra-smooth $CrTe_2$ surface and the *in-situ* integration process in the ultra-high vacuum environment result in the formation of a sharp hetero-interface, and the energy dispersive X-ray (EDX) spectroscopy images ascertain the uniform element distribution without observable inter-layer diffusion (Supplementary Section 3). After sample growth, standard 2D cross-bar device arrays were fabricated on the 2-inch $Bi_2Te_3$/$CrTe_2$ wafers (**Fig. 4b**), and the relevant $CrTe_2$-thickness-dependent current-driven magnetization switching results are presented in **Fig. 4c**. Guided by the schematic diagram in **Fig. 4a**, the presence of a constant in-plane field $B_x$ = +0.09 T sets up the initial magnetization state of the sample in the (+*x*, +*z*) quadrant, and the applied DC current $I_{DC}$ along the ±*x*-direction dictates the effective spin-orbit field $B_{SO} = \lambda_{SO} \cdot I_{TI} \cdot \sigma \times M$ (where $\lambda_{SO}$ is the coefficient characterizing the SOC strength, $I_{TI}$ is the charge current component conducting through the $Bi_2Te_3$ channel, and $\sigma = \mp\sigma_y \cdot \hat{y}$ is the electron spin accumulated at the $Bi_2Te_3$/$CrTe_2$ interface). Consequently, as the DC current is successively scanned from +25 to −25 mA, the measured Hall resistances $R_{xy}$ (blue circles in **Fig. 4c**) all retain at constant values until the positive-to-negative transitions occur in the large negative $I_{DC}$ domains (*i.e.,* the parallel (+$B_x$, +$I_{DC}$) configuration stabilizes the magnetic moment $M_z$ along the +z-axis, whereas the negative DC bias could trigger the magnetization switching as long as the reversed $B_{SO}$ overcomes the intrinsic magnetic anisotropy field $B_K$). On the other hand, the application of an opposite $B_x = -0.09$ T changes the initial condition of $M$, and the observation of the clockwise $R_{xy}$-$I_{DC}$ hysteresis loops (red triangles in **Fig. 4c**) are in accordance with the same damping-like SOT chirality, hence confirming the deterministic switching scenario.

Most intriguingly, the ($B_x$-fixed, $I_{DC}$-dependent) anomalous Hall data in **Fig. 4c** discloses a dramatic reduction of the threshold switching current level $I_{SW}$ when the top CrTe$_2$ layer thickness $d$ decreases from 21 ML (17.5 mA) to 5 ML (7.5 mA) at $T = 120$ K. Equivalently, it means that the dynamic power dissipation $P_{SW}$ can be reduced by up to 54% with appropriate bilayer structural engineering (**Fig. 4d**). To appreciate such a positive $I_{SW}$-$d$ correlation, it is recalled that the reduced dimensionality of the CrTe$_2$ film would not only weaken its perpendicular magnetic anisotropy (**Fig. 3**), but also make the thinner FM layer more insulating so that a larger portion of charge current is regulated inside the Bi$_2$Te$_3$ channel. Governed by the SOT operational principle that the modulation of the CrTe$_2$ FM order mainly relies on the competition/balancing between the torques exerted by the effective fields[33, 36], it can therefore be concluded that a lower DC current can meet the magnetization switching requirement of $|\gamma \cdot M \times B_{SO}| > \tau_K$ (where $\gamma$ is the gyromagnetic ratio, and $\tau_K$ is the torque exerted by PMA) thanks to the attenuated $B_K$ and higher $I_{TI}/I_{DC}$ ratio in the Bi$_2$Te$_3$/CrTe$_2$ (5 ML)-based SOT device. By further subtracting the shunting current via the parallel two-channel model (see Supplementary Section 4), the critical current density available for the charge-to-spin conversion is found to stay at $J_{TI, c} = 2.64 \times 10^6$ A·cm$^{-2}$ as long as the bottom CrTe$_2$ layer exceeds 9 ML, while it drops to $1.93 \times 10^6$ A·cm$^{-2}$ for the $d = 5$ ML case, again highlighting the unique advantage of vdW FMs in the 2D limit. Finally, we have evaluated the reliability of the SOT devices fabricated on the same Bi$_2$Te$_3$/CrTe$_2$ (5 ML) wafer. As summarized in **Fig. 4e**, the current-induced $R_{xy}$ loops for devices randomly picked from five different regions essentially coincide with each other, and the device-to-device variation of $I_{SW}$ is no more than 2.5%. Furthermore, the transition between the high- and low-resistance states in response to the alternating positive/negative input current pulses always maintain the identical square-wave-like contour without any distortion after $5 \times 10^4$ writing cycles (**Fig. 4f**), thus implying a high endurance of the tested SOT device.

**Conclusion**

In conclusion, we have achieved the epitaxial growth of uniform 1$T$-CrTe$_2$ films on a wafer-scale by MBE, and established a vdW integration strategy that allows for the *in-situ* construction of Bi$_2$Te$_3$/CrTe$_2$ heterostructures. Endorsed by the co-existence of the intrinsic PMA, atomically-sharp interfaces, and strong SOC, we showcase the compelling SOT device performance of this simple bilayer stack with lower switching power than the heavy metal-based FM multi-layer systems (green bars in **Fig. 4g**[37-40]). Moreover, its tunable vdW ferromagnetism makes CrTe$_2$ a competitive building block to directly pair with topological quantum materials without invoking an additional PMA-assisted layer (orange bars in **Fig. 4g**[34, 35, 40-43]). The design rule inaugurated in our work may expedite the search for new 2D vdW FMs-based heterostructures, which further enables large-scale vdW material synthesis and feasible spintronic device applications.

**Methods**

*Sample Fabrication and Structural Characterizations*: Wafer-scale CrTe$_2$ thin films and Bi$_2$Te$_3$/CrTe$_2$ heterostructures were grown on 2-inch Al$_2$O$_3$(0001) substrates by MBE in a vacuum of $1 \times 10^{-10}$ mbar. Prior to the sample growth, the Al$_2$O$_3$ substrate was pre-annealed at 600 °C in order to remove any absorbed contamination. During the MBE growth, the growth temperature for CrTe$_2$ and Bi$_2$Te$_3$ was kept at 200 °C, and the sample manipulator was constantly rotated to ensure uniformity; high-purity Cr and Bi atoms were evaporated from standard Knudsen cells, while the Te atoms were evaporated by a thermal cracker cell. A beam flux monitor was used to measure the flux ratio, and RHEED was used to monitor the real-time growth process. After sample growth, slices of $CrTe_2$ with different crystal orientations were milled out using the FIB technique (TESCAN LYRA3 FIB-SEM, TESCAN, Czech Republic), and a probe aberration-corrected scanning transmission electron microscopy (Cs-STEM, Themis Z G2 300, FEI, USA) was employed to resolve the crystal structure. The STEM was equipped with energy-dispersive X-ray analysis (EDX, Bruker Super-X, Bruker, USA), allowing for the determination of the Cr-to-Te stoichiometric ratio. Additionally, X-ray diffraction and reflectivity were used to confirm the lattice constant and to calibrate the thickness.

***Device Fabrication and Transport Measurements***: The $CrTe_2$ thin films were etched into a six-probe Hall bar geometry with typical channel size of $2 \times 1$ mm$^2$. The electrodes were made by welding small pieces of indium onto the contact areas. The magneto-transport measurements were performed in a $He^4$ refrigerator (Oxford Teslatron PT system). Multiple lock-in amplifiers and Keithley source meters (with an excitation AC current amplitude of $I = 1$ µA) were applied to the sample, and the temperature, magnetic field, and lock-in frequency served as the experimental variables. For the SOT-driven magnetization switching measurements, the cross-bar patterns of the $CrTe_2/Bi_2Te_3$ heterostructures were prepared by a standard photo-lithography process, and the Ti (15 nm)/Au (150 nm) electrodes were defined by e-beam evaporation. After device fabrication, a 2-ms writing current pulse was applied by Keithley 6221 while another 2-ms reading current pulse was applied thereafter to measure the $R_{xy}$ value by Keithley 2182.

***XAS and XMCD measurements:*** Element-specific X-ray absorption spectroscopy (XAS) and X-ray magnetic circular dichroism (XMCD) measurements were performed on beamline I06 at the Diamond Light Source (Oxfordshire, UK), allowing for the probing of the magnetic ground state of the $CrTe_2$ layers. The base temperature was 2 K and a magnetic field of ± 6 T was applied along the surface normal direction (out-of-plane). The spectroscopic data was acquired in total-electron-yield detection mode which has an 1/e sampling depth of ~5 nm[44]. The XAS spectra were recorded at the Cr $L_{2,3}$ edges between 560 and 595 eV. The XMCD results were obtained by taking the difference between two XAS spectra with the helicity vector of the circularly polarized X-rays parallel and antiparallel to the magnetic field, respectively.

***Density Functional Theory Calculations:*** The *ab initio* calculations were performed within the framework of density functional theory (DFT) as implemented in the Vienna ab initio simulation package (VASP)[45], with the exchange-correlation functional considered in the generalized gradient approximation of the Perdew-Burke-Ernzerhof[46] method. A plane-wave basis set was used with a kinetic energy cutoff of 500 eV. The *k*-meshes of 11×11×1 for the monolayers and slabs (including 4, 5, 7, and 8 ML) and 11×11×7 for the bulk $CrTe_2$ were applied. To account for the correlation effect of the transition metal element Cr, the GGA+*U* functional with varied *U* values for the *d*-orbitals of Cr were adopted. The spin-orbital coupling was considered self-consistently in this work. The intrinsic anomalous Hall conductivity were evaluated based on the maximally localized Wannier functions[47] as obtained through the VASP2WANNIER90[48] interfaces in a non-self-consistent calculation, and calculated on *k*-meshes of 1001×1001×1 for the monolayer, 601×601×1 for the slabs, and 301×301×301 for the bulk by the WannierTools package[49].

**Acknowledgments**

We thank the helpful discussion with Dr. Yilan Jiang. This work is sponsored by the National Key R&D Program of China under the contract number 2017YFB0305400, National Natural Science Foundation of China (Grant No. 61874172, 92164104 and 11904230), and the Major Project of Shanghai Municipal Science and Technology (Grant No. 2018SHZDZX02), the Shanghai Engineering Research Center of Energy Efficient and Custom AI IC, and the Shanghaitech Quantum Device and Soft Matter Nano-fabrication Labs (SMN180827). X.F.K acknowledges the support from the Merck POC program and the Shanghai Rising-Star program (Grant No. 21QA1406000). Q.Y acknowledges the support from the Shanghai Sailing Program (Grant No. 19YF1433200).

**Author Contributions**

X.F.K., and Q.Y. conceived and supervised the study. X.Q.L. and Q.Y. grew the samples and performed the X-ray measurements. X.Q.L. conducted the transport measurements. X.Q.L. and Q.Y. analyzed the transport and characterization data. L.G. and Y.X. conducted the TEM measurements and theoretical calculations, respectively. D.B., G.v.d.L., and T.H. performed the XAS/XMCD measurements. X.Q.L., Y.X., Q.Y., and X.F.K. wrote the manuscript. All authors discussed the results and commented on the manuscript.

**Competing interests**

The authors declare no competing interests.

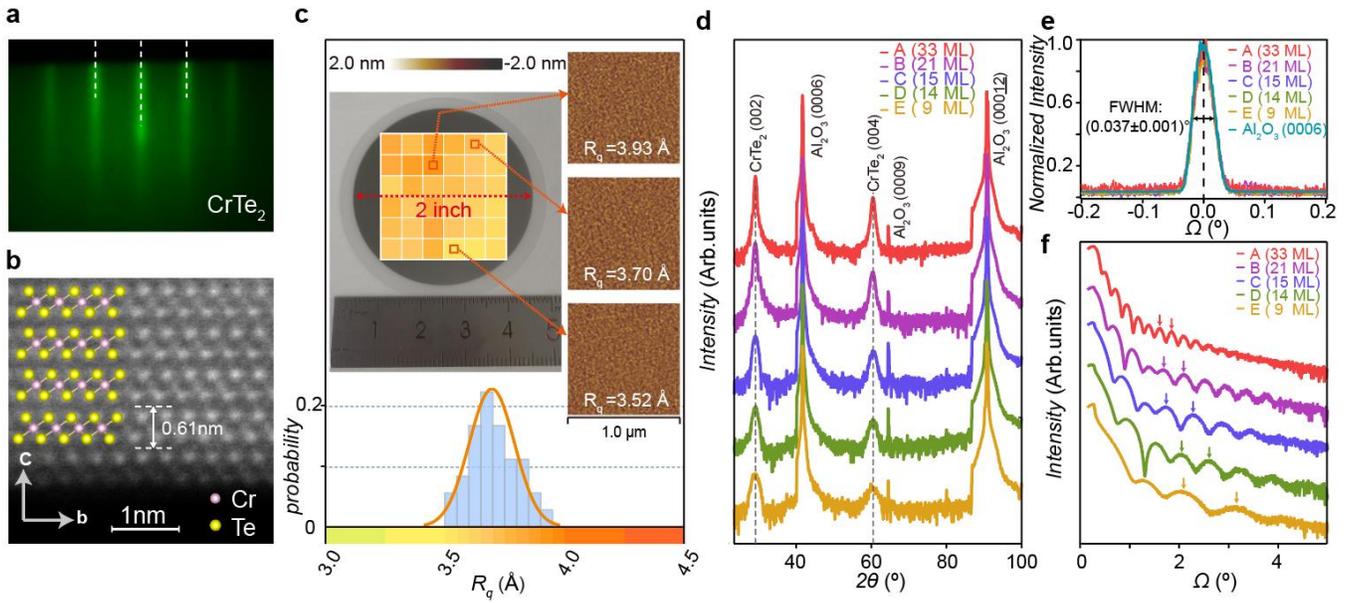

**Fig. 1. Epitaxial growth of wafer-scale 1*T*-CrTe$_2$ thin films. a,** *in-situ* RHEED pattern of the CrTe$_2$ film grown on the Al$_2$O$_3$ substrate. The spacing between the two first-order streaks (dashed white lines) is used to deduce the in-plane lattice constant. **b,** Cross-sectional High-resolution STEM image of the MBE-grown CrTe$_2$ sample. The atomic arrangement is consistent with the crystal structure of 1*T*-CrTe$_2$. The Cr and Te atoms are labeled as pink and yellow spheres, respectively. **c,** AFM images were taken at 36 different positions across the 2-inch 9 ML CrTe$_2$ wafer, and the statistical histogram of the measured root-mean-square roughness $R_q$ confirms the homogeneous surface morphology. **d,** The XRD spectra of CrTe$_2$ films ranging from 9 to 33 ML. **e,** Identical rocking curves of the CrTe$_2$ (002) film peaks with the same FWHM value as the Al$_2$O$_3$ substrate highlight the very high crystalline quality of the CrTe$_2$ films. **f,** XRR data of the thickness-dependent CrTe$_2$ films with oscillatory Kiessig fringes. The spacing between neighboring interference peaks (labeled as arrows) is inversely correlated to the film thickness.

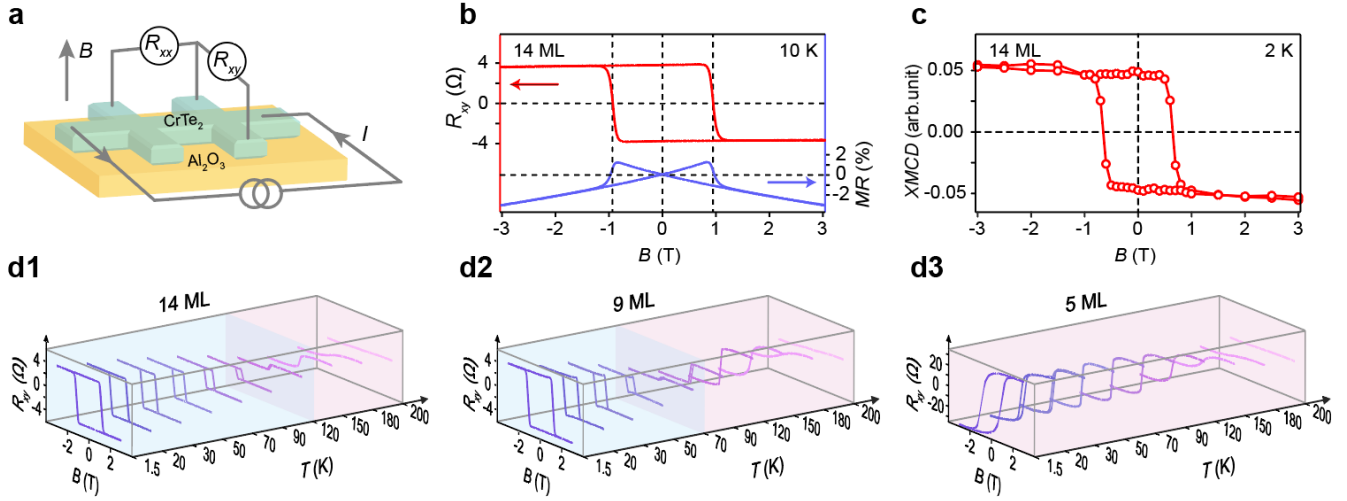

**Fig. 2. Thickness- and temperature-driven anomalous Hall effect in the MBE-grown CrTe$_2$ thin films. a,** Schematic of a six-probe Hall bar and the setup of the magneto-transport measurement. **b,** The anomalous Hall resistance ($R_{xy}$) and magnetoresistance (MR) curves of the 14 ML CrTe$_2$ film at $T = 10$ K. **c,** The square-shaped XMCD hysteresis loop taken at the Cr $L_3$ edge manifests the robust long-range magnetic moment with perpendicular magnetic anisotropy. **d1-d3**, Temperature-dependent AHE results in three CrTe$_2$ samples with film thickness of 14 ML (**d1**), 9 ML (**d2**), and 5 ML (**d3**), respectively. The AHE polarity of the bulk-type CrTe$_2$ samples is reversed in the high-temperature region, whereas such negative-to-positive transition is absent in the ultra-thin case.

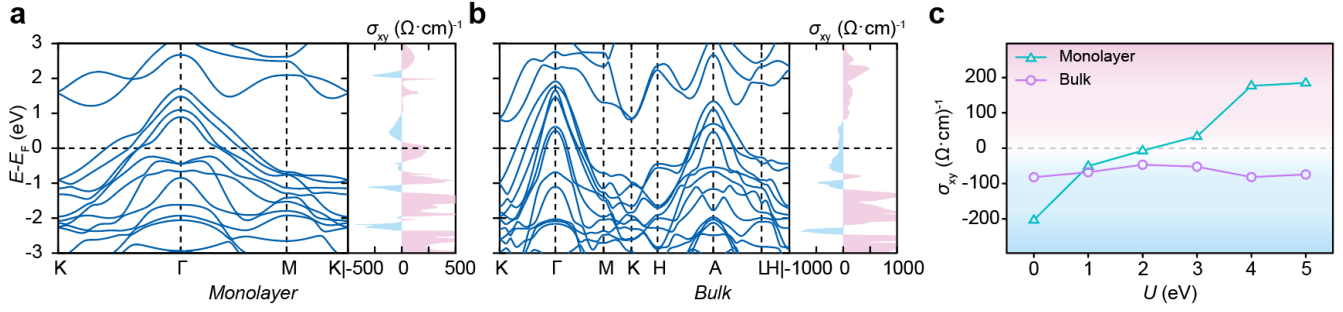

**Fig. 3. Capturing the dimensionality effect in the vdW FM CrTe$_2$ system by density-functional theory calculations. a-b**, Electronic band structure (left panel) and corresponding intrinsic anomalous Hall conductivity $\sigma_{xy}$ (right panel) of monolayer (**a**) and bulk-form 1$T$-CrTe$_2$ (**b**) with the onsite Coulomb potential of $U = 4$ eV. **c,** Evolution of the anomalous Hall conductivity with respect to the Coulomb potential. The negative $\sigma_{xy}$ of the bulk CrTe$_2$ shows little variation, while the orbital hybridization in the 2D limit modifies the Berry curvature and triggers the AHE polarity switching in the monolayer case.

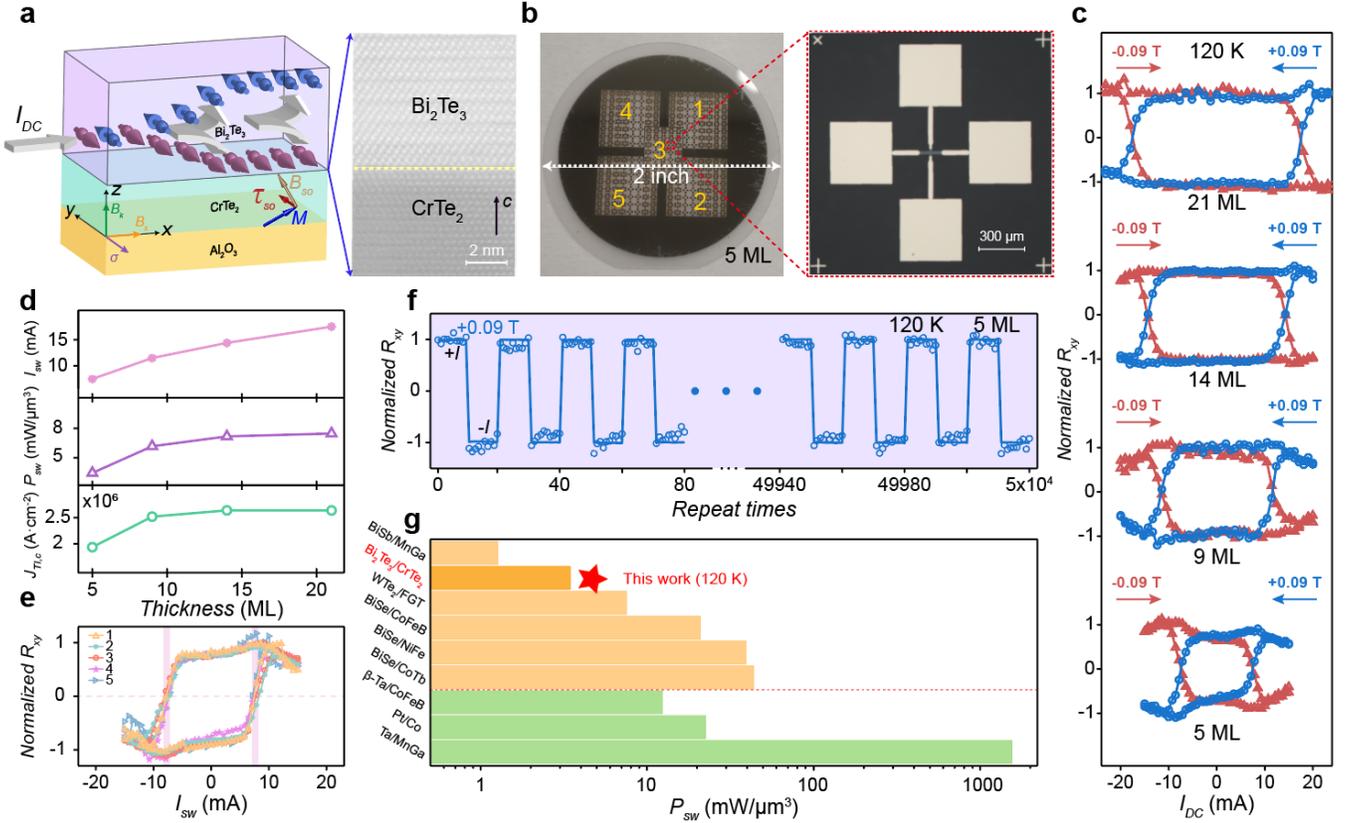

**Fig. 4. Tunable SOT-driven magnetization switching in Bi₂Te₃/CrTe₂ heterostructures. a,** Schematic of the SOT mechanism in the Bi₂Te₃/CrTe₂ bilayer stack. Right panel: the HR-STEM image discloses an atomically sharp Bi₂Te₃/CrTe₂ hetero-interface which warrants efficient spin injection. **b,** Optical image of the patterned cross-bar SOT device arrays on the 2-inch wafer. **c,** Current-induced magnetization switching in four Bi₂Te₃/CrTe₂ samples with different CrTe₂ layer thicknesses of 21, 14, 9, and 5 ML (from top to bottom). **d,** Summary of the critical switching current $I_{SW}$, dynamic power $P_{SW}$, and threshold current density $J_{TI,c}$ as a function of the CrTe₂ thickness. **e,** Highly-repeatable switching loops measured from five randomly-selected devices of the SOT device array. **f,** Endurance test of the Bi₂Te₃/CrTe₂(5 ML) device. The square-wave read-out signal does not show any distortion even after 5 × 10⁴ write/read cycles. **g,** Benchmark the performance of the Bi₂Te₃/CrTe₂-based SOT device with other magnetic bilayer systems.